\documentclass{ws-p8-50x6-00}
\usepackage{graphicx,overcite}
\newcommand{\alt}{\mathrel{\raisebox{-.6ex}{$\stackrel{\textstyle<}{\sim}$}}}
\newcommand{\agt}{\mathrel{\raisebox{-.6ex}{$\stackrel{\textstyle>}{\sim}$}}}

\begin{document}
\title{Mini-review on Collider Signatures for Extra Dimensions}

\author{Kingman Cheung}

\address{Department of Physics, University of California, Davis, CA 95616 USA
\\E-mail: cheung@gluon.ucdavis.edu}

\maketitle

\abstracts{
In this talk, I briefly review collider signatures for two models of
extra dimensions.  The first one was proposed by Arkani-Hamed {\it et
al.} that gravity is free to propagate in extra dimensions of very
large size ($\alt 1$ mm).  Collider signatures for this model can be
divided into two types: (i) emission of real gravitons into extra
dimensions, and (ii) exchanges of virtual gravitons.  The second model
was proposed by Pomarol {\it et al.} and Dienes {\it et al.} that the
SM gauge bosons are allowed to propagate in extra dimensions.
Collider signatures for the second model are due to the existence of
the Kaluza-Klein (KK) states of $\gamma$, $W$, $Z$, and $g$ bosons.
}

\section{Introduction}

Recent advances in string theory have revolutionized particle phenomenology.
Namely, the previously unreachable Planck, string, and grand unification 
scales ($M_{\rm Pl}$, $M_{\rm st}$, and $M_{\rm GUT}$, respectively) can 
be brought down to a TeV range through the existence of extra dimensions.
There have been a number of ideas that can bring either the Planck, GUT, 
or string scale down to TeV region. 
One expects the low energy phenomenology of these new ideas or models
can be tested at current and future collider experiments.

In this talk, we review collider signatures 
for the following two models of extra dimensions: (i) 
the one proposed by Arkani-Hamed, Dimopoulos, and Dvali \cite{ADD}, in
which the standard model (SM) particles live on a 3-brane while the gravity
is free to propagate in extra dimensions of very large size ($\alt 1$ mm).
This model was motivated by the fact that the effective Planck scale is 
brought down to TeV to solve the hierarchy problem.
Collider signatures for this model can be divided into two types:
(a) emission of real gravitons into extra dimensions and thus gives
rise to missing energy signals, and (b) exchanges of virtual 
gravitons that frequently lead to enhancement of 
production of SM particles.
(ii) The model proposed by Dienes {\it et al.} \cite{keith} and Pomarol 
{\it et al.} \cite{pomarol}
that the SM gauge bosons are allowed to propagate in extra dimensions 
(whereas the gravity effect here is negligible.)  This model has the merit of
unifying the gauge couplings at a scale much lower than the usual GUT scale.
Collider signatures for this model are due to
the existence of the Kaluza-Klein (KK) states of $\gamma$, $W$, $Z$, and 
$g$ bosons.  

\section{Model of Arkani-Hamed, Dimopoulos, and Dvali}

This model was first proposed to solve the hierarchy problem by requiring
the compactified dimensions to be of very large size, $\alt 1$ mm.
While the SM particles live on a D3-brane, the gravity
is free to propagate in extra dimensions.
Using Gauss law, the effective Planck scale $M_S$ is related to the 
four-dimensional Planck scale $M_{\rm Pl}$ ($10^{19}$ GeV) by
\[
M^2_{\rm Pl} \sim M_S^{n+2} \; R^n \;,
\]
where $n$ is the number of extra (compactified) dimensions and $R$ is the
size of the compactified dimensions. Assuming that the effective Planck scale
$M_S$ is in the TeV range,  it gives a very large $R$ of
the size of a solar system for $n=1$, which is obviously ruled out by
experiments. However, for all $n\ge 2$ the expected $R$ is less than 
$1$~mm, and therefore do not contradict existing gravitational experiments.

With the SM particles residing on the brane and the graviton freely
propagating in extra dimensions, 
in the 4D-point of view a graviton in extra dimensions is equivalent
to a tower of infinite number of KK states with masses 
$M_k = 2\pi k/R\;\; (k=0,1,2,...,\infty)$.
The couplings of SM particles to each of these KK states is still 
of order $1/M_{\rm Pl}$, but
the overall coupling is, however, obtained by summing over all the KK
states, and therefore scales as $1/M_S$. Since the $M_S$ is in the TeV
range, the effective gravitational interaction is as strong as the 
electroweak interaction and thus gives rise to many testable 
consequences in both accelerator and non-accelerator experiments.

A large number of phenomenological studies \cite{astro,giu,shrock,mira,han,
hewett,rizzo1,desh,grae,me1,rizzo2,atwood1,balazs,shiu,lee1,rizzo3,davo1,yosh,
lee2,he1,rain,he2,pila,das,sila1,davo2,moha,das-ray,ioan,sila2,cheng,gupta,
me2,bour,atwood2,me3,greg,eboli,mathews1,mathews2,mathews3,jack,lee3,gho,
atwood4,atwood3,donc,ghosh,chou,zhang,duane}
in this area have 
appeared.  The collider signatures can be divided into two categories:
(i) real emission of gravitons into extra dimensions, giving rise to
missing energy signal, and (ii) virtual exchange of gravitons in addition to
exchanges of SM gauge bosons, giving rise to enhancement or deviations
from the SM predictions.  We summarize these signals in the following.
Note that a stringent constraint
comes from astrophysical (SN1987A) and cosmological sources~\cite{astro}
and the lower bound on the effective Planck scale $M_S$ is 30--100 TeV for 
$n=2$.

\subsection{Real Emission of Gravitons}

Since  gravitons interact weakly with detectors, they 
will escape detection and causing missing energies.
Thus, the logical signal to search for 
would be the associated production of gravitons with other SM particles.  
At $e^+ e^-$ colliders, the best signals would be the associated production
of graviton with a $Z$ boson, a photon, or a fermion pair.

The production of graviton and photon at LEPII has been studied 
\cite{giu,mira,me1}.  The striking signature would be a single photon
with missing energy while the irreducible background comes from the 
process $e^+ e^- \to \gamma \nu_\ell \bar \nu_\ell,\; (\ell=e,\mu,\tau)$.
The cross section for the signal is given by \cite{giu,mira,me1}
\begin{eqnarray}
\frac{d\sigma}{d \cos\theta} &=& \frac{\pi \alpha G_N}{ 4 \left( 1- 
\frac{m^2}{s} \right )} \; \biggr[ (1+\cos^2\theta) \left( 1 +
(\frac{m^2}{s})^4 \right ) \nonumber \\
&+& \left( \frac{ 1-3\cos^2\theta +4\cos^4\theta}{1-\cos^2\theta} \right )
\, \frac{m^2}{s} \, \left( 1+ (\frac{m^2}{s})^2 \right ) +
6 \cos^2\theta (\frac{m^2}{s})^2 \biggr ] \;.
\end{eqnarray}
The signal cross section increases with the energy of collision while the
background is gradually decreasing after the $Z$-peak. At LEPII, if the
effective Planck scale $M_S$ is low enough deviations from the SM prediction
should be seen.  In a recent search by L3 \cite{L3-1}, the limit
$M_S\agt 1$ TeV for $n=2$.

The production of graviton with a $Z$ boson at LEPII \cite{me1}
gives a signature of a $Z$ boson, which decays
into a $q\bar q$ or $\ell\bar\ell$ pair, plus missing energy.  This is an
interesting process because LEPII already searched for the invisibly 
decaying Higgs boson in $ZH$ production, which has the same signature as 
the $ZG$ production.  The formulas for the signal cross section can be found
in Ref. \cite{me1}.  The irreducible background is $e^+ e^- \to Z\nu\bar\nu$.
Based on the existing data on the search for invisibly decaying Higgs, the
lower limit on $M_S$ is around 515 GeV.  A refined search by L3 \cite{L3-1}
gives a limit $M_S> 600$ GeV.

The associated production of graviton with a $f\bar f$ pair was studied at
$Z^0$ \cite{balazs} because of a large number of hadronic $Z$ decays.  The
signature would be a fermion pair with missing energy.  The background
comes from $Z\to f\bar f \nu \bar \nu$, which has a small BR of
$2\cdot 10^{-7}$.  The present data agrees with the SM prediction and is
able to constrain $M_S$ to be at least 0.4 TeV for $n=2$ \cite{balazs}.

Another exciting opportunity is the monojet plus missing energy production 
\cite{giu,mira} at hadronic colliders.

\subsection{Virtual Exchanges of gravitons}

There are numerous studies in collider signatures associated 
with virtual exchanges of KK-gravitons, including diphoton, 
diboson, and fermion-pair production.
In the following, we highlight some of these studies.

One of the most prominent channels is photon-photon scattering.
In the SM, photon-photon scattering only takes place via box diagrams of 
$W$ bosons and quarks so that it is loop-suppressed.  On the other hand,
in the ADD model photons can scatter via exchanges of gravitons in 
$s,t,u$-channels.  The scattering amplitude-squared is symmetric in $s,t,u$
and given by  \cite{me2}
$\overline{\sum} |{\cal M}|^2 = \frac{\kappa^4}{8} \; |D(s)|^2 \;
(s^4 + t^4 + u^4 )$.  
The differential cross section is given by
\begin{displaymath}
\frac{d\sigma (\gamma\gamma \to \gamma\gamma)}{d |\cos\theta| } =
\frac{\pi s^3}{M_S^8}\, {\cal F}^2 \, \biggr [
1 + \frac{1}{8}\,(1+ 6 \cos^2\theta + \cos^4 \theta ) \biggr ] \;,
\end{displaymath}
where ${\cal F}=2/(n-2)$ for $n>2$.
The signal cross section easily surpasses the SM background at around 
$\sqrt{s}=0.5$ TeV for $M_S=4$ TeV \cite{me2}.  
The polarized scattering has also been studied in 
Ref. \cite{rizzo3,davo1,chou}.
Another interesting channel is neutrino-photon scattering \cite{duane}.

Diphoton production \cite{giu,me2} and also $WW,ZZ$ production \cite{desh}
at $e^+ e^-$ colliders have been studied. The effect of TeV scale gravity 
on the angular distribution of diphoton production \cite{me2} is
\begin{equation}
\label{eeaa}
\frac{d\sigma (e^+ e^- \to \gamma \gamma)}{dz} =
\frac{2 \pi}{s}\; \Biggr(
\alpha  \sqrt{\frac{1+ z^2}{1- z^2}}
+ \frac{s^2}{8}\, \frac{\cal F}{M_S^4} \sqrt{1-z^4}
\Biggr )^2
\end{equation}
where $z=|\cos\theta_\gamma|$.
This effect is very similar to the conventional deviation from QED,
which is often parametrized by a $\Lambda$ as 
\[
\frac{d\sigma}{dz} = \frac{2\pi \alpha^2}{s}\; \frac{1+z^2}{1-z^2} \; \left(
1 \pm \frac{s^2}{2 \Lambda_\pm^4}\, (1 - z^2 ) \right )\;.
\]
We can immediately equate the above two expressions and arrive at
$\frac{M_S^4}{\cal F} = \frac{\Lambda_+^4}{2 \alpha}$.  The present limits
obtained by LEP Collaborations on 
$\Lambda_{\rm QED} \sim 262 - 345$ GeV, which is equivalent to 
$M_S \sim 0.7 - 1$ TeV for $n=4$ \cite{me2}.

Agashe and Deshpande \cite{desh} also studied $e^+ e^- \to \gamma\gamma, 
W^+ W^-,ZZ$ production and compared their sensitivities to TeV scale gravity. 
Interestingly, $ZZ$ production offers the highest fractional change of
cross section among $\gamma\gamma, WW,ZZ$ due to gravity effects.  
However, the $ZZ$ production rate is smaller than the other two.  Overall,
their sensitivities are similar.  A recent experimental search performed 
by L3 \cite{L3-1} found that the sensitivities of $\gamma\gamma,WW,ZZ$
are very similar and the combined limit is $M_S \agt 0.8$ TeV.

Diphoton production \cite{giu,me2,greg,eboli} is one of the best probes 
of TeV scale gravity at hadron colliders.  The angular distributions of the
subprocesses are given by
\begin{eqnarray}
\frac{d\sigma (q\bar q \to \gamma \gamma)}{d\cos\theta^*} &=&
\frac{1}{48\pi \hat s}\; \Biggr [
 e^2 Q_q^2 \,\sqrt{\frac{1+\cos^2 \theta^*}{1-\cos^2\theta^*}}
+
\frac{\pi \hat s^2}{2}\, \frac{\cal F}{M_S^4} \,\sqrt{1-\cos^4\theta^*} \,  
  \Biggr ]^2 \;, \\
\frac{d\sigma (gg \to \gamma \gamma)}{d\cos\theta^*} &=&
\frac{\pi \hat s^3}{512}\, \left( \frac{\cal F}{M_S^4}\right)^2 
\, (1+ 6 \cos^2\theta^* + \cos^4\theta^*) \;, 
\end{eqnarray}
where $\cos\theta^*$ is the scattering angle of the photon in the
center-of-mass frame of the incoming partons, and here $\cos\theta^*$ is from 
$-1$ to 1.
Based on the existing diphoton data from CDF \cite{cdf-1} that in 
$M_{\gamma\gamma}>150$ GeV 5 events are observed where $4.5\pm0.6$ are 
expected with a luminosity of 100 pb$^{-1}$, a limit of $M_S>0.9$ TeV 
\cite{me2} for $n=4$ was obtained.  The upcoming
CDF and D0 searches will easily overshadow this limit.

The general vector-boson scattering $VV\to VV$, where $V=\gamma,W,Z$ was
studied by Atwood {\it et al.} \cite{atwood2}.  The conclusion is that the 
effect of TeV scale gravity shows up at large invariant mass region.

Extra dimensions also affect
fermion-pair production at $e^+ e^-$ colliders and the corresponding crossing
channels, such as Drell-Yan production at the Tevatron and 
neutral-current (NC) deep-inelastic scattering (DIS) at HERA.
While they were individually studied in a number of publications
\cite{giu,hewett,rizzo1,gupta,mathews2}, a 
comphrensive analysis of all these data sets 
was performed in Ref. \cite{me3}. The combined limit on $M_S$ is
$M_S> 0.94$ TeV for $n=4$.  Bourilkov \cite{bour}, on the other hand, used the 
combined data of Bhabha scattering of the four LEP Collaborations and was
able to obtain a limit of $M_S>1.4$ TeV.
There are also a combined search in fermion-pair production, diphoton,
$WW$ and $ZZ$ pair production by L3 \cite{L3-1} that a limit of 
$M_S\agt 1$ TeV is established. 

Dijet and top-pair production \cite{mathews1,mathews3,atwood3,donc,ghosh}
at the Tevatron or other colliders should also be useful in obtaining
information on $M_S$, but systematics will likely reduce the usefulness 
of these channels.
Effects on precision variables \cite{grae,das-ray,zhang}
and effects on patterns of fermion or neutrino masses \cite{yosh,ioan,das,moha}
have also been studied.  Interactions with scalars or Higgs bosons have
been investigated in Ref. \cite{rizzo2,he1,he2,jack}.

In the following, we describe a few studies that test sensitivity to $M_S$ in
future experiments at hadronic and $e^+ e^-$ colliders.
Cheung and Landsberg \cite{greg} used double differential cross-sections,
$d^2\sigma/d M d \cos\theta^*$, of diphoton and Drell-Yan production
to constrain the effective Planck
scale $M_S$ in Run I and Run II at the Tevatron and at the LHC.  
The advantage of using double differential distributions is that the
differences in invariant mass and scattering angle between the SM 
and the gravity model can be contrasted simultaneously.
Furthermore, for a $2\to 2 $ process the invariant mass $M$ and the
central scattering angle $\cos\theta^*$ already span the entire phase
space.  We, therefore, do not need further optimization of cuts or
variables.  The resulting sensitivities to $M_S$ are shown in Table 
\ref{table-X}(a).

\begin{table}[t]
\caption{\small \label{table-X}
95\% C.L. sensitivity limits on $M_S\;(n=4)$ at (a) hadronic colliders
(Tevatron and LHC) and (b) $e^+ e^-$ colliders of 0.5 -- 1.5 TeV.}
\medskip
\centering
\begin{tabular}{|c|c|c|c|}
\hline
\multicolumn{4}{|c|}{(a) Hadronic Colliders} \\
\hline
\hline
Run I & Run IIa & Run IIb & LHC  \\
${\cal L}=0.13$ fb$^{-1}$ & 2 & 20 & 100 \\
\hline
1.3 & 1.9 & 2.6 & 9.9 \\
\hline
\end{tabular}
\begin{tabular}{|cc|ccc|cc|}
\hline
\multicolumn{7}{|c|}{(b) $e^+ e^-$ Colliders} \\
\hline
\hline
\multicolumn{2}{|c|}{$\sqrt{s}=0.5$ TeV} & 
\multicolumn{3}{c|}{1 TeV } & \multicolumn{2}{c|}{1.5 TeV } \\
${\cal L}=10$ fb$^{-1}$     & 50  & 10  & 50  & 100 & 100  & 200 \\
\hline
3.9 & 4.8 & 6.5 & 7.9 & 8.9 & 12.0 & 13.0   \\
\hline
\end{tabular}
\end{table}

The sensitivity reach at linear $e^+ e^-$ colliders was studied in a number of
work \cite{giu,hewett,rizzo1,desh,lee1,lee2,lee3}.  
Here I present a work \cite{me4} that uses diphoton, Bhabha
scattering, $\mu^+\mu^-$, $\tau^+\tau^-$, $q\bar q$ production and their
angular distributions.  The sensitivity limits on $M_S$ are obtained by a
combined maximum likelihood approach by adding the likelihoods of all
channels.  The Bhabha scattering turns out to be the dominant channel.  The
angular distribution for diphoton production is given in Eq. (\ref{eeaa})
and for fermion-pair production it is given by
\begin{eqnarray}
\frac{d \sigma(e^- e^+ \to f \bar f)}{dz} &=& 
\frac{N_f s }{128\pi} \Biggr \{ (1+z)^2 (|M_{LL}(s)|^2 + |M_{RR}(s)|^2)
+  (1-z)^2  \nonumber \\
&& \hspace{-0.6in} \times\; ( |M_{RL}(s)|^2 + |M_{LR}(s)|^2) 
+ \pi^2 s^2 ( 1-3z^2 + 4z^4) \eta^2  \nonumber \\
&& \hspace{-0.6in} -\; 8\pi e^2 Q_e Q_f z^3 \eta
+ \frac{8\pi e^2}{ s^2_\theta c^2_\theta} \frac{s}{s - M_Z^2} \left(
g_a^e g_a^f \frac{1-3z^2}{2} - g_v^e g_v^f z^3 \right) \eta \Biggr \}
      \nonumber \\
&& \hspace{-0.6in} +\; \frac{\delta_{e f} s}{128 \pi}
  \Biggr \{ (1+z)^2 (|M_{LL}(t)|^2 + |M_{RR}(t)|^2 
         + 2 M_{LL}(s) M_{LL}(t)  \nonumber \\
&& \hspace{-0.6in} +\; 2 M_{RR}(s) M_{RR}(t) )
 + 4 ( |M_{LR}(t)|^2 + |M_{RL}(t)|^2 ) 
+ \frac{\pi^2 s^2}{8}  \nonumber \\
&& \hspace{-0.6in} \times \;(121 + 228z +198z^2 + 84z^3 + 9z^4)\eta^2 
-  
\frac{\pi s}{2} \eta ( M_{LL}(t) + M_{RR}(t) \nonumber \\
&&\hspace{-0.6in} +\; M_{LL}(s) + M_{RR}(s) ) (1+z)^2
 (7+z)  + \pi s \eta (M_{LL}(t) + M_{RR}(t) )  \nonumber \\
&&\hspace{-0.6in} \times \; (1+z)^2 (1-2z)  
- 2 \pi s \eta ( M_{LR}(t) + M_{RL}(t) ) (5+3z)  \Biggr \} \nonumber
\end{eqnarray}
where $\eta={\cal F}/M_S^4$ and the reduced amplitudes are given by
\[
M^{ef}_{\alpha\beta}(s) = 
\frac{e^2 Q_e Q_f}{s} + \frac{e^2}{\sin^2\theta_{\rm w}
\cos^2 \theta_{\rm w}} \, \frac{g_\alpha^e g_\beta^f}{s - M_Z^2} \;,
\;\;\;\;\; \alpha, \beta= L,R  \;.
\]
The 95\% C.L. sensitivity limits on $M_S$ for $n=4$ are shown in Table 
\ref{table-X}(b).  The values for other $n$ can be easily obtained by 
scaling with ${\cal F}=2/(n-2)$ for $n>2$.  From the table we can see 
that a 0.5 TeV NLC has a reach at least double of that at RunIIa and 
still higher than that of RunIIb. The reach by the LHC is only slightly 
better than the 1 TeV NLC.

\section{Kaluza-Klein States of SM Gauge Bosons}

In another brane configuration, the SM particles reside
on a $p$-brane ($p=\delta+3 >3$) whereas the gravity is in the rest
$(6-p)$ dimensional bulk.  Within this $p$-brane the effect of gravity is
negligible compared to gauge interactions.  
Inside the $p$-brane the chiral fermions are restricted
to a 3-brane and the gauge bosons can propagate in the extra $\delta$ 
dimensions internal to the $p$-brane.  

Dienes {\it et al.} \cite{keith} considered in this model of extra dimensions
and showed that the gauge couplings can be unified at a scale lower than
the usual GUT scale, due to the power running of the couplings.  Therefore,
the effective GUT scale is lowered, which is in contrast to the ADD model 
that the Planck scale is lowered.

For collider phenomenology a specific model of this type was proposed by
Pomarol {\it et al.} \cite{pomarol}. It is a five-dimensional
model with the fifth dimension compactified on a $S^1/Z_2$.  In the 
four-dimensional point of view, a gauge boson $V$ that propagates in the
fifth dimension is equivalent to a tower of Kaluza-Klein states $V^{(n)}$ with
mass $M_n = n M_c$, where $M_c=1/R$ is the compactification scale and $R$ is
radius of the fifth dimension.  The resulting 4-D Lagrangian for 
charged-current and neutral-current interactions are, respectively, given by
\begin{eqnarray}
{\cal L}^{\rm CC} &=& \frac{g^2 v^2}{8} \biggr [ W_1^2 + \cos^2\beta
\sum_{n=1}^{\infty}( W_1^{(n)})^2 + 2\sqrt{2} \sin^2\beta W_1 
 \sum_{n=1}^{\infty} W_1^{(n)} + 2 \sin^2\beta \nonumber \\
&\times&\left( \sum_{n=1}^{\infty} W_1^{(n)} \right )^2 \biggr ]
+ \frac{1}{2} \sum_{n=1}^{\infty} n^2 M_c^2 (W_1^{(n)})^2 
 - {g} (W_1^\mu + \sqrt{2} \sum_{n=1}^{\infty} W_1^{(n)\mu} ) J_\mu^1 
\nonumber \\
 &+& (1 \to 2)  \label{cc} \\
{\cal L}^{\rm NC} &=& \frac{{g}v^2}{8 {c}_\theta^2} \biggr [ Z^2
+ \cos^2\beta \sum_{n=1}^{\infty}( Z^{(n)})^2 + 2\sqrt{2} \sin^2\beta Z 
 \sum_{n=1}^{\infty} Z^{(n)} + 2 \sin^2\beta \nonumber \\
&\times& \left (\sum_{n=1}^{\infty}  Z^{(n)} \right )^2
+ \frac{1}{2} \sum_{n=1}^{\infty} n^2 M_c^2 \biggr[ (Z^{(n)} )^2 + 
  (A^{(n)})^2 \biggr] \nonumber \\
&-& \frac{{e}}{ {s}_\theta {c}_\theta } \left (
 Z^\mu + \sqrt{2} \sum_{n=1}^{\infty} Z^{(n)\mu} \right ) J_\mu^Z 
- {e} \left (
 A^\mu + \sqrt{2} \sum_{n=1}^{\infty} A^{(n)\mu} \right ) J_\mu^{\rm em} 
\label{nc}
\end{eqnarray}

There are two important effects
of these KK states on collider experiments. (i) Since 
the KK states have the same quantum numbers as their corresponding gauge 
bosons, it gives rise to mixing effects between the zeroth (the SM gauge 
boson) and the $n$th-modes of $W$ and $Z$ bosons.  The zero mass of the photon
is protected by the U(1)$_{\rm EM}$ symmetry of the SM. 
(ii) If the energy is higher than the compactification scale $M_c$, real 
emissions or resonances of KK states of $\gamma,W,Z,g$ bosons can be 
observed, otherwise enhancement of cross sections may be possible.

In  Eqs. (\ref{cc}) and (\ref{nc}) the first few terms of each imply
mixings among $V,V^{(1)},V^{(2)}, ...$ ($V=W,Z$). 
The mixing between the SM gauge bosons with its Kaluza-Klein states
modifies electroweak observables (similar to the mixing between
the $Z$ and a $Z'$) via a series of mixing angles, 
which depend on the masses of $Z^{(n)}, n=0,1,...$ and the angle $\beta$.
The neutral boson at LEP is then the first mass eigenstate after 
mixing.  The couplings of the $Z^{(0)}$ to fermions are modified through 
the mixing angles.  The observables at LEPI can place strong constraints on the
mixing, and thus on the compactification scale $M_c$.
Similarly, the properties of the $W$ boson are also modified. 
The effects on electroweak precision measurements have been studied 
\cite{nath,rizzo,casa,Strumia,carone,del,CC}.  
Overall, the limit on $M_c$ using the precision data measurements is
$M_c \agt 3.3 - 3.8$ TeV.

The effects of Kaluza-Klein states of the SM gauge bosons 
also occur in high energy processes.  
If the available energy is higher than the compactification scale, real 
emissions or resonances of these Kaluza-Klein states can be observed.  However,
for the present collider energies and because the compactification scale is 
believed to be at least a few TeV, only indirect effects can be seen.

We summarize a study of high energy processes when $M_c$ is higher than the
present energy scale. The reduced amplitude for $q\bar q \to \ell^+ \ell^-$ or
$\ell^+ \ell^- \to q\bar q $ is given by
\begin{equation}
\label{mab}
M^{eq}_{\alpha\beta}(s) = e^2 \Biggr \{ \frac{Q_e Q_q}{s} + 
\frac{g_\alpha^e g_\beta^q}{s^2_\theta c^2_\theta} \;
\frac{1}{s - M_Z^2 } 
- \left( Q_e Q_q +  \frac{g_\alpha^e g_\beta^q}
     {s^2_\theta c^2_\theta } \right ) \;
 \frac{\pi^2}{3 M_c^2 } \; \Biggr \} \;.
\end{equation}
where $M^2_c \gg s, |t|, |u|$.
The above formula is applicable to hadronic and leptonic cross sections at
$e^+ e^-$ colliders and to Drell-Yan production at the Tevatron, and with
a crossing to DIS at HERA.  Similar expressions can be found \cite{me5}
for $W$ KK state exchanges and for dijet and $t\bar t$ production.
In a global fit to $\eta= \pi^2/3 M_c^2$, Cheung and Landsberg~\cite{me5}
include the following data sets: (i) LEPII hadronic and all leptonic
production cross sections and angular distributions, (ii) Drell-Yan 
production at the Tevatron, (iii) NC and CC DIS scattering cross sections
at HERA, (iv) Tevatron dijet production, and (v) Tevatron $t\bar t$ 
production.  The resulting 95\% C.L. limit on $M_c$ is 
\[
M_c > 4.4\; {\rm TeV} \;.
\]

\section{Conclusions}

Physics in extra dimensions and phenomenology are extremely rich with advance
in string theories because fundamental scales are now reachable within
future collider experiments.  We have briefly reviewed collider signatures
of two interesting scenarios.  The first model is due to the KK states of
gravitons while the second one is due to KK states of gauge bosons.
It turns out that diphoton, boson-pair, and fermion-pair production,
as well as precision data measurements are useful in probing these 
two models.

\section*{\bf Acknowledgments}
I thank Wai-yee Keung and Greg Landsberg for parts of the work 
presented here.  This research was supported in part by the 
DOE Grants No. DE-FG03-91ER40674.



\end{document}